\documentstyle[12pt]{article}

\catcode`\@=11
\def\marginnote#1{}
\newcount\hour
\newcount\minute
\newtoks\amorpm
\hour=\time\divide\hour by60
\minute=\time{\multiply\hour by60 \global\advance\minute by-\hour}
\edef\standardtime{{\ifnum\hour<12 \global\amorpm={am}%
        \else\global\amorpm={pm}\advance\hour by-12 \fi
        \ifnum\hour=0 \hour=12 \fi
        \number\hour:\ifnum\minute<10 0\fi\number\minute\the\amorpm}}
\edef\militarytime{\number\hour:\ifnum\minute<10 0\fi\number\minute}
\def\draftlabel#1{{\@bsphack\if@filesw {\let\thepage\relax
   \xdef\@gtempa{\write\@auxout{\string
      \newlabel{#1}{{\@currentlabel}{\thepage}}}}}\@gtempa
   \if@nobreak \ifvmode\nobreak\fi\fi\fi\@esphack}
        \gdef\@eqnlabel{#1}}
\def\@eqnlabel{}
\def\@vacuum{}
\def\draftmarginnote#1{\marginpar{\raggedright\scriptsize\tt#1}}
\def\draft{\oddsidemargin -.5truein
        \def\@oddfoot{\sl preliminary draft \hfil
        \rm\thepage\hfil\sl\today\quad\militarytime}
        \let\@evenfoot\@oddfoot \overfullrule 3pt
        \let\label=\draftlabel
        \let\marginnote=\draftmarginnote
   \def\@eqnnum{(\theequation)\rlap{\kern\marginparsep\tt\@eqnlabel}%
\global\let\@eqnlabel\@vacuum}  }

\def\preprint{\twocolumn\sloppy\flushbottom\parindent 1em
        \leftmargini 2em\leftmarginv .5em\leftmarginvi .5em
        \oddsidemargin -.5in    \evensidemargin -.5in
        \columnsep 15mm \footheight 0pt
        \textwidth 250mmin      \topmargin  -.4in
        \headheight 12pt \topskip .4in
        \textheight 175mm
        \footskip 0pt
        \def\@oddhead{\thepage\hfil\addtocounter{page}{1}\thepage}
        \let\@evenhead\@oddhead \def\@oddfoot{} \def\@evenfoot{} }

\def\titlepage{\@restonecolfalse\if@twocolumn\@restonecoltrue\onecolumn
     \else \newpage \fi \thispagestyle{empty}\c@page\z@ 
        \def\thefootnote{\fnsymbol{footnote}} }

\def\endtitlepage{\if@restonecol\twocolumn \else  \fi
        \def\thefootnote{\arabic{footnote}}
        \setcounter{footnote}{0}}  %\c@footnote\z@ }

\catcode`@=12
\relax
\def\bea{\begin{array}}
\def\bem{\begin{displaymath}}
\def\beq{\begin{equation}}

\def\eea{\end{array}}
\def\eem{\end{displaymath}}
\def\eeq{\end{equation}}
\def\NP#1#2#3{Nucl. Phys. {\bf #1} (19#2) #3}
\def\ov{\overline}

\def\PL#1#2#3{Phys. Lett. {\bf #1} (19#2) #3}

\def\s2w{\sin^2 \theta_W}

\relax
\def\crbig{\\\noalign{\vspace {3mm}}}

\def\S{\Sigma}

\def\hpsi{{\hat\psi}}

\def\be{\begin{equation}}
\def\ee{\end{equation}}
\def\ba{\begin{eqnarray}}
\def\ea{\end{eqnarray}}

\def\r{\rho}
\def\a{\alpha}

\def\b{\beta}

\def\g{\gamma}

\def\d{\delta}

\def\e{\epsilon}

\def\p{\pi}

\def\m{\mu}
\def\n{\nu}
\def\om{\omega}

\def\l{\lambda}

\def\s{\sigma} 
\def\S{\Sigma}

\def\qq{\qquad}

\def\IR{\relax{\rm I\kern-.18em R}}

\def \ha {{1\over 2}}

\def \ov {\over}

\def\inv{^{\raise.15ex\hbox{${\scriptscriptstyle -}$}\kern-.05em 1}}

\def\hi{{\hat i}}
\def\hj{{\hat j}}
\def\hk{{\hat k}}

\def\be{\begin{equation}}
\def\ee{\end{equation}}
\def\ba{\begin{eqnarray}}
\def\ea{\end{eqnarray}}

\def\a{\alpha}
\def\b{\beta}
\def\g{\gamma}

\def\d{{\rm d}}
\def\e{\epsilon}
\def\p{\psi}

\def\m{\mu}
\def\n{\nu}
\def\r{\rho}
\def\l{\lambda}

\def\s{\sigma}
\def\o{\omega}
\def\f{\phi}

\def\ks{{k \kern-.5em /}}
\def\es{{\e \kern-.4em /}}
\def\ds{{\partial \kern-.5em /}}
\def\Ds{{D \kern-.6em /}}

\def\R{{\cal R}}
\def\ph{\hat\psi}

\def\qq{\qquad}

\def \ha {{1\over 2}}
\def \ov {\over}
\def\inv{^{\raise.15ex\hbox{${\scriptscriptstyle -}$}\kern-.05em 1}}

\relax
\renewcommand{\theequation}{\thesection.\arabic{equation}}
\begin{document}
\topmargin-2.4cm
%\draft
%\preprint
%
%
%
%
\begin{titlepage}
\begin{flushright}
NEIP--01--011 \\
LPTENS-01/45\\
hep--th/0111274 \\
November 2001
\end{flushright}
\vskip 3.5cm

\begin{center}{\Large\bf
(Weak) $G_2$ holonomy from self-duality,\break
flux and supersymmetry}
\vskip 1.5cm
{\bf Adel Bilal$^{1,2}$, Jean-Pierre Derendinger$^1$ and
Konstadinos Sfetsos$^{1,3}$}
\vskip .1in
$^1$ Institut de Physique,
Universit\'e de Neuch\^atel \\
CH--2000 Neuch\^atel, Switzerland

\vskip.3cm
$^2$ Laboratoire de Physique Th\'eorique, Ecole Normale Sup\'erieure\\
24 rue Lhomond, 75231 Paris Cedex 05, France

\vskip.3cm
$^3$ Department of Engineering Sciences, University of Patras\\
26110 Patras, Greece\\

\end{center}
\vskip .5cm

\begin{center}
{\bf Abstract}
\end{center}
\begin{quote}
The aim of this paper is two-fold. First,
we provide a simple and pedagogical discussion 
of how compactifications of M-theory or 
supergravity preserving some four-dimensional 
supersymmetry naturally lead to reduced holonomy or its generalization, 
reduced {\it weak} holonomy. We relate the existence of a (conformal) 
Killing spinor to the existence of certain  closed and co-closed
$p$-forms, and to the metric being Ricci flat or Einstein.\\
Then, for seven-dimensional manifolds, we show that 
octonionic self-duality conditions on the spin connection are  equivalent 
to $G_2$ holonomy and certain generalized self-duality conditions to 
weak $G_2$ holonomy. The latter lift to self-duality conditions for 
cohomogeneity-one ${\rm spin}(7)$ metrics. To illustrate the power of
this approach, we present several examples where the self-duality 
condition largely simplifies the derivation of a $G_2$ or weak $G_2$ 
metric.
\end{quote}

\noindent\rule{8cm}{.1mm}

\noindent
{\small
e-mail: {\tt adel.bilal,
jean-pierre.derendinger, konstadinos.sfetsos@unine.ch}}
\end{titlepage}
\setcounter{footnote}{0}
\setcounter{page}{0}
\setlength{\baselineskip}{.7cm}
\newpage
%
% BODY
%

\section{Introduction\label{Intro}}

Compactifications to four dimensions of M-theory or string theory, in
their simplest setting (no background flux or gauge fields turned on),
require the compactification manifold $K$ to admit at least one covariantly
constant spinor in order to preserve some four-dimensional supersymmetry. 
The existence
of a covariantly constant spinor in turn implies that the manifold $K$ does
have reduced holonomy: ${\rm SU}(3)$ instead of  ${\rm SU}(4)\simeq 
{\rm SO}(6)$ in string theory and $G_2$ instead of ${\rm SO}(7)$  in
M-theory/eleven-dimensional supergravity. It also implies that $K$ is Ricci
flat and hence the metric field equations are automatically satisfied in the
compact directions. For these reasons, 6-dimensional manifolds with 
${\rm SU}(3)$ holonomy (Calabi-Yau spaces) have been abundantly studied over
the last decade and a half. Similarly, 
seven-dimensional manifolds with $G_2$
holonomy have attracted much attention more recently. Mathematically quite
similar to the latter are also eight-dimensional manifolds of 
${\rm spin}(7)$ 
holonomy, and both cases can often be studied in parallel, but the real
physical interest of course is in seven-dimensional manifolds.

One might ask what happens in a less simple setting when background fields
are turned on. Let us now concentrate on M-theory, or on its low-energy 
limit, eleven-dimensional
supergravity. Physically, the background fields induce a non-vanishing
energy-momentum tensor and the metric is no longer Ricci flat. The next
best thing we can hope for is that $K$ is an Einstein space, {\it i.e.}
$R_{ij}\sim g_{ij}$. Conditions for unbroken four-dimensional supersymmetry
now are the existence of a conformal Killing spinor $\eta$ on $K$
satisfying $D_i\eta \sim \g_i \eta$. This indeed implies that $K$ is an
Einstein manifold. Being no longer Ricci flat, it cannot have $G_2$
holonomy. However, we will see that the appropriate notion in this setting
is that of {\it weak $G_2$ holonomy}, a concept originally introduced by
Gray \cite{Gray}. While $G_2$ holonomy is equivalent to the existence of a
3-form $\f$ obeying $\d\f=0$ and $\d *\f=0$, weak $G_2$ holonomy is
equivalent to the existence of a 3-form $\f$ obeying $\d\f=\l *\f$ and then
also $\d *\f=0$. Any weak $G_2$ holonomy manifold is an Einstein space with
$R_{ij}\sim \l^2 g_{ij}$, and we will see how the condition of weak $G_2$
holonomy is related to the conformal Killing spinor equation $D_i\eta \sim
\l \g_i \eta$. In this sense, one may view $G_2$ holonomy as the 
$\l\to 0$ limit of weak $G_2$ holonomy.

Many metrics of $G_2$ holonomy and some metrics of weak $G_2$ holonomy have
been constructed over the past (for some examples, see {\it e.g.} 
\cite{Bryant,GPP,Cvetic2,BGGG,Cvetic1,GUSPA}), not always without 
considerable
effort. In the present paper, we will show how a simple self-duality
condition for the  spin connection $\o^{ab}$ allows for a quick
and easy proof of all desired properties. The basic ingredient to define
self-duality in seven dimensions is the $G_2$ invariant 4-index tensor which
is dual to the octonionic structure constants. (Recall that $G_2$ is the
automorphism group of the octonion algebra.) A similar method in eight
dimensions was first employed some time ago in \cite{Bakas} to find an
Eguchi-Hanson type metric of ${\rm spin}(7)$ holonomy which had previously 
been found in \cite{GPP} by solving second order differential equations. 
Also refs. \cite{FK,Brecher} use a self-duality in seven dimensions to get 
$G_2$ holonomy. In a way this is
natural, since using
self-duality of $\o^{ab}$ is a well-known technique in four dimensions to
generate self-dual and hence Ricci-flat curvatures. 
Actually, it was first realised in \cite{AOL} that imposing an 
octonionic self-duality condition on the curvature $R^{ab}$ in 
eight dimensions leads to 7 relations between the 28 components 
and hence implies that the holonomy group must be ${\rm spin}(7)$. 
By doing trivial dimensional reduction, the same argument implied that 
self-dual curvature in seven dimensions must correspond to $G_2$ holonomy, 
in six dimensions to ${\rm SU}(3)$ holonomy etc.\footnote{ 
The relation between self-dual curvature 
and self-dual spin connection was also briefly mentioned in ref. \cite{AOL}.
A different approach was outlined in ref. \cite{YO} and used 
in ref. \cite{KN}.}

In four dimensions the
Eguchi-Hanson metric \cite{EH}
does follow from imposing self-duality of $\o^{ab}$
within a certain ansatz for the metric, but the Atiyah-Hitchin metric 
\cite{AH} does
not obviously seem to fall into this class. However, it can be obtained 
from adding an appropriate inhomogeneous term
to the self-duality condition. This motivated us to study generalized 
self-duality conditions for the 
seven-dimensional spin connection, allowing also
for certain inhomogeneous terms. While it was known \cite{FK,Brecher} that
in seven dimensions self-duality implies $G_2$
holonomy, we show that some appropriately generalized self-duality condition
with an inhomogeneous term implies {\it weak} $G_2$ holonomy. 
Although self-duality
of $\o^{ab}$ is a ``gauge"-dependent condition, in four dimensions it is
known (see {\it e.g.} \cite{EGH}) that a self-dual curvature always follows 
from an $\o^{ab}$ that is
self-dual up to a local Lorentz transformation.\footnote{It 
can actually be shown that in the case of the Atiyah-Hitchin 
metric a ``non-canonical" choice of the
4-bein, related by some local Lorentz transformation to the obvious one, 
is equivalent to the inhomogeneous term mentioned above. Such an 
equivalence will obviously not be possible between $G_2$ and weak $G_2$ 
holonomy.}
We prove the analogous 
statement for $G_2$ and weak $G_2$: for any manifold of (weak) $G_2$ 
holonomy, the spin connection is (generalized) self-dual, up to a 
local ${\rm SO}(7)$ transformation. For most of the
$G_2$ and weak $G_2$ holonomy metrics that are known in the 
literature, the spin connection
actually does satisfy the self-duality criterion without any need
to perform a local Lorentz transformation. A standard example for weak $G_2$
holonomy is the metric based on the Aloff-Walach spaces
$N(k,l)$ (cosets ${\rm SU}(3)/{\rm U}(1)$) and, with the obvious choice,
its spin connection satisfies our generalized self-duality
condition. (This corresponds to an analogous statement made in 
\cite{KY} but can also be easily checked directly.)

With hindsight, most of these results could have been expected since 
weak $G_2$
holonomy metrics can be generated from cohomogeneity-one ${\rm spin}(7)$
metrics \cite{Hitchin}, and indeed our inhomogeneously modified self-duality
condition for weak $G_2$ holonomy is related to the standard self-duality of 
${\rm spin}(7)$. On the other hand, we are not aware of a proof 
that weak $G_2$ holonomy implies the existence of a (generalized) 
self-dual $\o^{ab}$. Ref. \cite{AOL} {\it e.g.} only condiders trivial 
reductions from 8 to 7 dimensions resulting in Ricci flat $G_2$, 
not weak $G_2$, manifolds. 
Thus, although pieces of information are scattered
throughout the literature, to the best of our knowledge, they have not been
all linked together. At the risk of repeating one or the other fact 
well-known to the experts, we try to give a self-contained presentation 
in which self-duality relations play a central r\^ole.

The outline of the paper is the following. In the next section, we first
review some well-known facts on covariantly constant spinors and reduced
holonomy. In section 3, we introduce the self-duality condition for
$\o^{ab}$ and show equivalence with holonomy contained in $G_2$, implying
self-duality of the curvature and Ricci flatness.
To underscore the power of this approach, we show how to get in a few
lines the differential equations for the general six-function ansatz for a 
$G_2$ metric obtained in
\cite{BGGG}. In section 4, we turn to the modified
self-duality condition and show how it implies weak $G_2$ holonomy, a
modified version of self-duality of the curvature, and the Ricci tensor being
Einstein. We also prove the converse, namely that for any weak $G_2$ 
holonomy manifold we can always find a spin connection satisfying the 
generalized self-duality condition. We discuss the relation with 
cohomogeneity-one ${\rm spin}(7)$
metrics which provides non-trivial examples of weak $G_2$ holonomy
satisfying the generalized self-duality conditions.
In section 5, we return to the physics of M-theory, resp. eleven-dimensional
supergravity compactified on a seven-manifold and in
particular rederive that unbroken four-dimensional supersymmetry 
only allows a four-form flux in space-time, namely 
$F_{\m\n\r\s}=f \e_{\m\n\r\s}$, while the
internal components $F_{ijkl}$ all must vanish \cite{BEdWN, CR}.
The seven-manifold then has weak $G_2$ holonomy, and
the flux strength $f$
turns out to be proportional to the weak $G_2$ holonomy parameter 
$\l$, {\it i.e.} $\d\f \sim f *\f$. This is the precise statement how weak 
$G_2$ holonomy 
is related to supersymmetric compactification.
Finally, we conclude in section 6.
Since the octonionic structure constants and their duals play a prominent
r\^ole in all our paper, an appendix is devoted to their properties and
collects useful identities.

\section{(Conformal) Killing spinors and reduced (weak) holonomy}
\setcounter{equation}{0}

\subsection{Killing spinors, reduced holonomy and Ricci flatness}

We begin by reviewing a few well-known facts about reduced holonomy. 
As discussed in the introduction, in the absence of background fields 
besides the metric tensor, a compactification of supergravity will preserve 
some supersymmetry if the compactification manifold $K$ admits a Killing 
spinor, {\it i.e.} a covariantly constant (non-zero) spinor $\eta$. 
This is reviewed in some detail in section 5. So assume 
\be\label{di}
D_i \eta =0 \, , \quad\quad i=1,\ldots, {\rm dim}\,\, K \,.
\ee
Applying another $D_j$ and antisymmetrising, this leads to
\be\label{dii}
R_{abij}\g^{ab}\eta=0 \ .
\ee
Our conventions are that $a,b,c,\ldots$ are flat and $i,j,k,\ldots$ 
are curved indices on $K$. The viel-bein one-forms will be denoted 
$e^a=e^a_i \d x^i$.
But $R_{abij}\g^{ab}$ are the generators of the holonomy group and 
eq. (\ref{dii}) precisely tells us that the holonomy group cannot be the 
maximal $SO({\rm dim} K)$ but must be a subgroup of it: holonomy has been
reduced. 

In seven dimensions, the generic spinor transforms as representation 
${\bf 8}$ of $SO(7)$. If one spinor solves eq. (\ref{di}), one expects
(at least) the decomposition ${\bf8} \rightarrow {\bf7}+{\bf1}$ under 
the holonomy group and this is the decomposition under 
$G_2 \subset {\rm SO}(7)$. Conversely, 
if the holonomy is $G_2 \subset {\rm SO}(7)$, then the spinor splits as 
${\bf 8}\to {\bf 7}+{\bf 1}$ and the ${\bf 1}$  corresponds to a 
covariantly constant (actually constant and $G_2$ invariant) spinor. 
In general, the existence of a covariantly constant spinor always 
implies reduced holonomy, but the converse is not always true. Actually 
the converse can only be true if the space is Ricci flat, since (\ref{di}) 
always implies Ricci flatness as we now recall.

The (torsion-free) curvature $R^{ab}=\ha R^{ab}_{\ \ cd} e^c\wedge e^d$ 
satisfies
$R_{abcd}=R_{cdab}=-R_{abdc}$ as well as $R_{a[bcd]}=0$ and the Ricci 
tensor is $\R_{ij}=R^a_{\ iaj}$. Now multiply (\ref{dii}) by $\g^j$ to 
conclude that 
\be\label{diii}
\R_{ij}\g^j\eta=0 ,
\ee
which in turn implies 
\be\label{div}
\R_{ij}=0 .
\ee

Another criterion for reduced holonomy more often used in the 
mathematical literature is the existence of certain closed $p$-forms, 
{\it e.g.} a closed 4-form $\Phi$ for ${\rm spin}(7)$ and a closed 3-form 
$\f$ with its dual 4-form $*\f$ also closed for $G_2$ holonomy. This 
can be easily understood from the existence of a covariantly constant 
spinor. It is well-known that from a (commuting) spinor one can make $p$-forms as 
\be\label{dv}
\f_{(p)}={(-i)^p\over p!}\, \overline\eta \g_{i_1 \ldots i_p}\eta
\, \d x^{i_1} \wedge \ldots\wedge \d x^{i_p} \, .
\ee
In the absence of torsion, 
\be\label{dvi}
\d \f_{(p)}=0
\ee
is equivalent to 
$D_{[j} \left( \overline\eta \g_{i_1 \ldots i_p]}\eta \right) =0$
and  we see that (\ref{di}) 
implies eq. (\ref{dvi}). To have equivalence, both equations must contain 
the same number of equations. Consider the example of $G_2$. Then 
$\eta$ is an 8-component (real) spinor and one can only make a 
non-vanishing 3-form $\f\equiv \f_{(3)}$ and of course
its dual 4-form $*\f$. Imposing then
\be\label{dvii}
G_2\ : \quad \d \f=0\quad {\rm and} \quad \d *\f=0
\ee
corresponds to $\pmatrix{7\cr 4\cr} + \pmatrix{ 7\cr 5\cr}= 56$
conditions. But $D_i\eta=0$ also corresponds to $7\times 8=56$ 
conditions. 
Hence, conditions  (\ref{dvii}) are equivalent to (\ref{di}), 
and thus a necessary and sufficient condition for the holonomy to be 
contained in $G_2$. Actually, if we have more than one Killing spinor 
the holonomy may be further reduced. Consider the example of the 
seven-dimensional product manifold $S^1\times CY$ where the holonomy 
is ${\rm SU}(3) \subset G_2$. Then 
${\bf 8}\to ({\bf 3} + {\bf \overline 3} + {\bf 1})+ {\bf 1}$. If 
we want to exclude such cases in order to have exactly $G_2$ holonomy, 
not some subgroup of it, we should require the existence of 
one and only one (real) Killing spinor.

\subsection{Conformal Killing spinors, reduced weak holonomy and 
Einstein spaces}

Similarly, it is easy to see that in general
\be\label{dviii}
D_j \eta = i \tilde\l\, \g_j \eta
\ee
implies that $d\phi_{(p)}$ is proportional to $\phi_{(p+1)}$.
Specifically, for $SO(7)$ spinors the relation is
\be\label{dix}
\d \f_{(p)}=[1-(-1)^p](p+1) \,  \tilde\l \, \f_{(p+1)} \, ,
\qquad\qquad p=0,\ldots,6,
\ee
(where $\phi_{(1)}=
\phi_{(2)}=0$ and $\phi_{(7-p)}=*\phi_{(p)}$).
With $p=3$, $\ \f_{(3)}\equiv \f$ and $\f_{(4)} =*\f_{(3)}\equiv *\f$, 
this reads
\be\label{dx}
\d\phi =\l *\f \quad {\rm with}\quad \l=8\tilde\l \,,
\ee
and we see that (\ref{dviii}) implies weak $G_2$ holonomy. Since (\ref{dx}) 
obviously implies $\d *\f=0$, the count of independent conditions is
unchanged and (\ref{dx}) and (\ref{dviii}) are actually equivalent.

Acting on (\ref{dviii}) with $D_i$ and antisymmetrising, one gets
\be\label{dxi}
R^{ab}_{\ \ ij} \g_{ab} \eta= 8 \tilde\l^2\, \g_{ij} \eta \, ,
\ee
so that now $\eta$ is not invariant under the holonomy group but 
transform in a particularly simple form. If one multiplies (\ref{dxi}) 
by $\g^j$ one gets $\R_{ij}\g^j\eta = 4 (d-1) \tilde\l^2 \g_i\eta$ where 
$d$ is the dimension of the manifold $K$, and one concludes that $K$ 
is an Einstein space with
\be\label{dxii}
\R_{ij}=4 (d-1) \tilde\l^2\, g_{ij} \, .
\ee
We repeat the relevant formula for the case of present interest:
\be\label{dxiii}
\begin{array}{rrcl}
{\rm weak}\ G_2 \,\, :& \quad \quad\quad 
D_j\eta &=&{i\over 8}\l\, \g_j\eta \,,\crbig
& \d\f &=& \l *\f \,, \crbig
&\R_{ij} &=&{3\over 8} \l^2\, g_{ij} \,. 
\end{array}
\ee

\section{$G_2$ holonomy from self-duality, and vice versa}
\setcounter{equation}{0}

In this section, we exclusively consider seven-dimensional manifolds. 
As before, 
$a,b, \ldots$ $= 1, \ldots 7$ are flat indices while 
$i,j,\ldots = 1, \ldots 7$ are curved ones.

\subsection{Self-duality equations}

An antisymmetric tensor $A^{ab}$ transforms as the adjoint representation 
${\bf 21}$ of ${\rm SO}(7)$. Under 
the embedding of $G_2$ into ${\rm SO}(7)$, we have
\be\label{ti}
{\bf 21}\to {\bf 14}+{\bf 7} \, ,
\ee
${\bf 14}$ being the adjoint of $G_2$ and ${\bf 7}$ the fundamental 
representation. We will call $A_+^{ab}$, resp. $A_-^{ab}$ the 
part transforming as the ${\bf 14}$, resp. ${\bf 7}$ of $G_2$. 
The decomposition can be performed with projectors using the 
$G_2$ invariant 4-index tensor $\ph_{abcd}$ which is the dual of the 
structure constants $\p_{abc}$ of the imaginary octonions, $G_2$ being 
the automorphism group of the latter. Both $\p_{abc}$ and $\ph_{abcd}$, 
$a,b,c,d=1,\ldots 7$, are completely antisymmetric and only take the 
values $\pm 1$ and 0. 
They are given in the appendix together with 
many useful identities (see also {\it e.g.} references \cite{GG,GT,dWN}). 
We only repeat here the most important 
for us:
\ba\label{tii}
\ph_{abfg}\ph_{fgcd}
&=&-2\ph_{abcd}+4\delta_{ac}\delta_{bd}-4\delta_{ad}\delta_{bc} \, ,
\\
\label{tiii}
\ph_{abde}\p_{dec}&=&-4 \p_{abc} \,, 
\\
\label{tiiia}
\p_{acd} \p_{bcd} &=& 6\delta_{ab} \, .
\ea
Consider then the orthogonal projectors\footnote{Sub- and superscripts 
have the same significance.} \cite{dWN}
\beq
\label{defproj}
({\cal P}_{\bf14})_{ab}^{\ \ cd} = {2\over3}\left(\delta_{ab}^{cd}
+{1\over4} \hat\psi_{ab}\,^{cd}\right) , \qquad\qquad
({\cal P}_{\bf7})_{ab}^{\ \ cd} = {1\over3}\left(\delta_{ab}^{cd}
- {1\over2}\hat\psi_{ab}\,^{cd}\right) ,
\eeq
with
$\delta_{ab}^{cd} = {1\over2}(\delta_a^c\delta_b^d - \delta_a^d\delta_b^c)$. Since $({\cal P}_{\bf14})_{ab}^{\ \ cd}\psi_{cde} = 0$
and $({\cal P}_{\bf7})_{ab}^{\ \ cd}\psi_{cde} = \psi_{abe}$, 
the representation
{\bf7} can be written as $\psi_{abc}\zeta^c$ and the representation
{\bf14} can be obtained by imposing the seven conditions 
$\psi_{abc}A^{ab}=0$.
The decomposition reads then
\beq
\label{tiv}
\begin{array}{lll}
A^{ab} = A_+^{ab} + A_-^{ab}, &&
\crbig
A_+^{ab} = {2\over3}\bigl(A^{ab} + {1\over4}\hpsi^{abcd}A^{cd}\bigr), \qquad
&\psi^{abc}A_+^{bc}=0,
\qquad &(\hbox{{\bf14} of }G_2),
\crbig
A_-^{ab} = {1\over3}\bigl(A^{ab} - {1\over2}\hpsi^{abcd}A^{cd}\bigr)
\,\,\equiv\,\, \psi^{abc}\zeta^c, \qquad
&\zeta^c = {1\over6}\psi^{cab}A^{ab},
\qquad &(\hbox{{\bf7} of }G_2).
\end{array}
\eeq
In other words, the two $G_2$ irreducible components can be obtained by
imposing on the adjoint of $SO(7)$ the $G_2$ invariant self-duality
conditions
\beq
\label{tv}
\begin{array}{rcll}
A_+^{ab} &=& {1\over2}\hpsi_{abcd}A_+^{cd}, \qquad &(\hbox{{\bf14} of }G_2),
\crbig
A_-^{ab} &=& -{1\over4}\hpsi_{abcd}A_-^{cd}, \qquad &(\hbox{{\bf7} of }G_2),
\end{array}
\eeq
which respectively eliminate the {\bf7} or the {\bf14} component of $A^{ab}$.
Alternatively, one might write out explicitly the relations (\ref{tv}) 
to see that they give 7 linear relations between the $A_+^{ab}$ 
(leaving 14 independent quantities) and 14 linear relations between 
the $A_-^{ab}$ (leaving 7). The relations (\ref{tv}) are analogous to the 
self-duality equations $F^{\m\n}=\pm \ha \e^{\m\n\r\s} F^{\r\s}$ in four 
dimensions realising  the decomposition ${\bf 6}\to {\bf 3}+ {\bf 3}$, 
except that now the coefficients $\pm \ha$ are replaced by $\ha$ and 
$-{1\over 4}$. We will refer to $A_+^{ab}$ as the self-dual part and 
to $A_-^{ab}$ as the anti self-dual part of $A^{ab}$. Since the
projectors (\ref{defproj}) are orthogonal and complementary, one has, 
of course, for any 
antisymmetric $A^{ab}$ and $B^{ab}$
\be\label{tvii}
A^{ab} B^{ab} = A_+^{ab} B_+^{ab} + A_-^{ab} B_-^{ab}  \, .
\ee
Each term is invariant under $G_2$ while the sum is invariant under $SO(7)$.

It is now evident that the 14 generators $G^{ab}$ 
of the $G_2$ algebra are the projection with ${\cal P}_{\bf14}$ of 
the 21 generators $J^{ab}$ of ${\rm SO}(7)$ \cite{dWN}, 
\be\label{tviii}
G^{ab}= J_+^{ab} \, .
\ee
The component invariant under $G_2$ of an $SO(7)$ spinor will then verify
\be\label{tix}
\g_+^{ab}\eta \equiv {2\over 3} 
\left( \g^{ab} +{1\over 4} \ph_{abcd} \g^{cd} \right) \eta = 0 \, .
\ee
With the explicit representation of the $\g$-matrices in 
terms of the octonionic structure constants given in the appendix, 
eq. (\ref{tix}) implies $\eta_\a=0$ for $\a=1,\ldots, 7$ and $\eta_8$ 
is the $G_2$ singlet direction. In particular, $G_2$ invariant spinors 
always exist (in seven dimensions) provided there is no global 
obstruction to the existence of spinors.

\subsection{Self-dual spin connection, Killing spinors and \\
$G_2$ holonomy}

Assume now that the spin connection one form
$\o^{ab}=\o^{ab}_j \d x^j$ satisfies any of the three equivalent statements
\be\label{tx}
\begin{array}{rcl} 
\psi_{abc}\,\o^{bc}&=& 0 \,, \cr
\o_-^{ab}&=& 0 \,, \cr
\o^{ab}&=&\ha \ph_{abcd}\, \o^{cd} \ , 
\end{array}
\ee
{\it i.e.} $\o^{ab}=\o^{ab}_+$.
The covariant derivative of a spinor then is
\be\label{txi}
D_j\eta= \left( \partial_j +{1\over 4} \o^{ab}_j \g^{ab} \right)\eta 
= \left( \partial_j +{1\over 4} \o^{ab}_j \g_+^{ab} \right)\eta \,.
\ee
With (\ref{tix}),  
one deduces that a $G_2$ invariant ($\g_+^{ab}\eta=0$)
and constant ($\partial_j\eta=0$) 
spinor is covariantly constant with respect to $\o^{ab}$. As mentioned 
above, in the representation (\ref{gamma}) of the $\g$-matrices, this is 
simply $\eta_\a={\rm const} \times \delta_{\a 8}$. Thus any of the three 
equations (\ref{tx}) implies the existence of (at least) 
one Killing spinor 
and hence the holonomy is contained in $G_2$.

The converse is also true: $G_2$ holonomy implies 
that we can find a self-dual spin connection $\o^{ab}=\o_+^{ab}$. 
This means that, 
given a $G_2$ manifold with an $\o^{ab}$, it must be possible to make a 
local ${\rm SO}(7)$ transformation such that the 
new connection satisfies $\o_-^{ab}=0$.
To show this, suppose that we have $G_2$ holonomy. Then there exists a 
single Killing spinor $\eta$ with $D_j \eta=0$. 
This means that $\eta(x)$ and $\eta(x+\d x)$ only differ by an 
${\rm SO}(7)$ rotation with parameters $\o_i^{ab}\d x^i$ and
the ``length" $\overline\eta \eta=\eta^\tau\eta$ remains unchanged. 
But we may at any point rotate $\eta$ so that it only has an eighth 
component\footnote{On a single spinor direction, the action of $SO(7)$ 
is due to the generators in the coset $SO(7)/G_2\sim SO(8)/SO(7)$. 
This can also be verified using our choice of gamma 
matrices and the results given in the appendix.}.
As the ``length" does not change 
as we vary $x^j$, we see that $\eta_\a=c\, \delta_{\a 8}$ with constant 
$c$, so that $\partial_j\eta=0$. Then $D_j\eta=0$ reduces to
\be\label{txiia}
0=D_j\eta = {1\over 4} \o_j^{ab} \g^{ab} \eta \ .
\ee
Since  $\eta_\a=c\, \delta_{\a 8}$,  we see from the appendix that
$\bigl(\g^{ab} \eta\bigr)_\a = c\, \p_{ab\a}$, 
so that (\ref{txiia}) reads $\o_j^{ab} \p_{ab\a}=0$. But by (\ref{tiv}) 
this is the same as saying that $\o_{-j}^{ab}=0$, which we wanted to 
show. This result is independent of the choice of a particular 
representation of $\g$-matrices and we conclude that

\noindent
$\bullet$
{\it The self-duality (\ref{tx}) of the spin connection 
$\o^{ab}$ implies 
$G_2$ holonomy (or holonomy contained within $G_2$)
and conversely, every $G_2$ holonomy metric can be derived 
from a self-dual connection satisfying (\ref{tx}).}
 
By the general arguments of section 2 we then know that 
$\d\f=0,\ \d *\f=0$ and $\R_{ij}=0$, but it will be instructive 
to rederive these results directly from the self-duality (\ref{tx}).
The relevant 3-form is (in the following the wedge product of forms 
will be implicitly understood)
\be\label{txii} 
\f={1\over 3!} \p_{abc}\, e^a e^b e^c \, , \quad \qquad
*\f={1\over 4!} \ph_{abcd}\, e^a e^b e^c e^d \,,
\ee
where $e^a$ is the 7-bein one-form related to $\o^{ab}$ by
\be\label{txiii}
\d e^a + \o^{ab} e^b =0 \, .
\ee
Note that using the explicit representation of the $\g$-matrices 
given in the appendix, as well as $\eta_\a = \delta_{\a 8}$ it is 
easy to see that (\ref{txii}) exactly coincides with the general 
definition (\ref{dv}). Using the self-duality (\ref{tx}) and  
the identity (\ref{ident3}) of the appendix, one has
\be\label{txiv}
\d\f = -{1\over 2} \p_{abc} e^a e^b \o^{cd} e^d 
= -{1\over 4} \p_{abc}  \ph_{cdef} e^a e^b \o^{ef} e^d 
= \p_{ade}  e^a e^d \o^{eb} e^b =-2\, \d \f \,.
\ee
Similarly, using (\ref{tx}) and the identity (\ref{ident4})
leads to 
\ba\label{txv}
\d *\f&=& -{1\over 6} \ph_{abcd} \o^{ae}  e^e e^b e^c  e^d 
= -{1\over 12} \ph_{aefg} \ph_{abcd}  \o^{fg} e^e e^b e^c e^d 
\nonumber\\
&=& -{1\over 3} \ph_{fbcd} \o^{fe} e^e e^b e^c  e^d = - 2\,  \d *\f  \, .
\ea
We then obtain
\be\label{tvi}
\d \f =0 \,, \quad\qquad \d *\f =0 \,, 
\ee
as required for $G_2$ holonomy.

Next, we show that self-duality of $\o^{ab}$, eq. (\ref{tx}), implies 
self-duality of the curvature 2-form 
$R^{ab}=\d \o^{ab}+\o^{ac}\o^{cb} \equiv \ha R^{ab}_{\ \ cd} e^c e^d$. 
For the $\d \o^{ab}$ part this is obvious. For the remaining part we 
have, using again (\ref{tx}) and the identity (\ref{ident4})
\be\label{txvii}
\begin{array}{rcl}
{1\over 2} \ph_{abcd} \o^{ce}\o^{ed} 
&=&{1\over 4} \ph_{abcd} \ph_{edfg} \o^{ce}\o^{fg}\crbig
&=&-{1\over 2} \ph_{abfe} \o^{fc}\o^{ce}
+{1\over 4} \ph_{aefg}\o^{be}\o^{fg} -{1\over 4} \ph_{befg} \o^{ae}\o^{fg}
+ \o^{ac}\o^{cb} \crbig
&=&-{1\over 2} \ph_{abcd} \o^{ce}\o^{ed} + 2 \o^{ac}\o^{cb} \,,
\end{array}
\ee
as desired. Hence
\be\label{txviii}
R^{ab}={1\over 2} \ph_{abcd} R^{cd} \, ,
\ee
which, as usual, immediately implies Ricci flatness:
\be\label{txix}
\R_{ab}=R_{acbc}=\ha \ph_{acde} R_{debc} = \ha \ph_{acde} R_{b[cde]}=0 \, .
\ee
Note that self-duality of the curvature is the most direct statement that 
the generators of the holonomy group are in $G_2$ rather than ${\rm SO}(7)$.

\subsection{A rather general example}

To illustrate the power of the self-duality equations, we consider an ansatz
for a
$G_2$ holonomy metric proposed in \cite{BGGG} 
and also independently in \cite{Cvetic1}. These authors 
have written a rather general cohomogeneity-one seven-dimensional 
metric ansatz with an ${\rm SU}(2) \times {\rm SU}(2) \times {\bf Z}_2$ 
symmetry acting on the manifold. It depends on six arbitrary functions of 
one coordinate $r$\ :
\beq
\d s^2 = \d r^2 + \sum_{i=1}^3 a_i^2(r) (\s_i-\S_i)^2  + 
\sum_{i=1}^3 b_i^2(r) (\s_i+\S_i)^2  \ ,
\eeq
where 
\beq
\d\s_i =-\ha \e_{ijk} \s_j \wedge \s_k\ ,\qq
\d\S_i =-\ha \e_{ijk} \S_j \wedge \S_k\ 
\eeq
are the left-invariant one forms for each of the  ${\rm SU}(2)$, and the
${\bf Z}_2$ acts by interchange $\s_i \leftrightarrow \S_i$. Note, that 
in this 
subsection $i,j,k$ are flat indices and run over 1, 2 and 3 only 
(contrary to our usual conventions!).
The obvious (though not unique) choice  of seven-bein is
\be
e^i  =   a_i (\s_i - \S_i)\ , \  i=1,2,3 , \qq
e^\hi  =   b_i (\s_i + \S_i)\ , \  \hi = i + 3\ , \qq
e^7   =   \d r\ .
\ee
Then
\ba
\d e^i & = & {\dot a_i\ov a_i} e^7 \wedge e^i - \ha {a_i\ov a_j b_k} 
\e_{ijk} e^j \wedge e^\hk\ ,
\nonumber\\
\d e^\hi & = &  {\dot b_i\ov b_i} e^7 \wedge e^\hi - {b_i\ov 4}
\e_{ijk} \left({e^j \wedge e^k\ov a_j a_k} 
+ {e^\hj \wedge e^\hk \ov b_j b_k}\right) \, ,
\ea 
from which one reads the spin connection:
\ba
&& \om^{i7}= {\dot a_i\ov a_i} e^i \ , \qq 
\om^{\hi 7}= {\dot b_i\ov b_i} e^\hi \ , \qq 
\om^{i\hi}=0 \ {\rm (no\ sum\ over\ } i)\ ,
\nonumber\\
&& \om^{ij} = {1\ov 4} \e_{ijk} \left({b_k\ov a_i a_j} - {a_i\ov a_j b_k} 
- {a_j\ov a_i b_k} \right ) e^\hk \ ,
\nonumber\\
&& \om^{\hi \hj} = {1\ov 4} \e_{ijk} \left({b_k\ov b_i b_j} 
- {b_i\ov b_j b_k} - {b_j\ov b_k b_i} \right ) e^\hk \ ,
\nonumber\\
&& \om^{i\hj} = {1\ov 4} \e_{ijk} \left({a_k\ov a_i b_j} 
- {a_i\ov a_k b_j} - {b_j\ov a_i a_k} \right ) e^k \ , \qq
\om^{\hi j} = - \om^{j \hi} \ .
\ea
Now impose the self-duality equations (\ref{tx}). The most convenient 
form to use is $\p_{abc}\o^{bc}=0$, since this directly only gives the 
7 linearly independent equations. Using eq. (\ref{split}), they can 
be written as
\ba\label{splitselfdual}
0&=& \p_{iab}\o^{ab}
=\e_{ijk} ( \om^{jk}-\om^{\hj\hk}) -2 \om^{7\hi} \ , 
\nonumber\\
0&=& \p_{\hi ab}\o^{ab}
= -\e_{ijk} ( \om^{\hj k} + \om^{j\hk}) + 2 \om^{7i} 
= 2 ( \om^{7i} -\e_{ijk} \om^{\hj k} )\ , 
\nonumber\\
0&=&\p_{7ab}\om^{ab} = 2 \om^{i\hi} \ .
\ea
The last equation is automatically satisfied while the  other six
immediately give the differential equations 
($\dot a_i = \ldots, \dot b_i=\ldots$) of ref. \cite{BGGG}.

Similarly, one may consider other cohomogeneity-one metrics with 
principal orbits given by a coset, {\it e.g.} ${\rm SU}(3)/{\rm U}(1)^2$, 
and impose self-duality to get first order differential equations 
for the coefficient functions.

\section{Weak $G_2$ holonomy and generalized self-duality}
\setcounter{equation}{0}

We now extend the results of the previous section to the case of weak 
holonomy. The self-duality condition on the spin connection is
generalized by the addition of a particular inhomogeneous term, 
transforming in the ${\bf 7}$ of $G_2$. We then show how this 
is equivalent to the existence of a conformal Killing spinor and 
hence to weak $G_2$ holonomy. We again rederive directly from the 
generalized self-duality that $\d \f=\l *\f$ and that the metric is 
Einstein, as it should. Finally we discuss the relations to and 
implications for ${\rm spin}(7)$ holonomy metrics.

\subsection{generalized self-duality and weak $G_2$ holonomy}

We do not any longer assume that the Levi-Civita spin connection transforms 
as a pure {\bf14} of $G_2$ but allow for a non-zero piece proportional 
to $\psi_{abc}\o^{bc}$ transforming as a {\bf7}. The non-trivial 
condition we require is that this piece is proportional to the seven-bein 
$e^a$. We can write the condition 
in the following three equivalent ways  (remember that 
$\psi_{abc}\o_+^{bc}=0$ is trivially true):
\beq\label{w1}
\begin{array}{rcl}
\psi_{abc}\o^{bc} &=& -{\lambda\over2}\, e^a \,, \crbig
\o_-^{ab} &=& -{\lambda\over12}\, \psi_{abc} e^c \,, \crbig
\o^{ab} &=& {1\over2}\hat\psi_{abcd}\, \o^{cd} 
- {\lambda\over4}\, \psi_{abc}\, e^c\,.
\end{array}
\eeq
Eqs. (\ref{w1}) now imply weak $G_2$ holonomy. To verify this, we 
recall from section 2 that we only 
need to show that there exists a spinor $\eta$ which satisfies
\beq
\label{w2}
D_j\eta = i{\lambda\over8}\, \gamma_j\eta \,.
\eeq
We pick a $G_2$ invariant constant spinor $\eta$ obeying (\ref{tix}).
As discussed before, such a spinor always exists. Then
\beq
\label{w3}
D_j\eta = {1\over4}\, \o_{-j}^{ab}\, \gamma_-^{ab}\eta 
= -{\lambda\over48\, }e_j^c\, \psi_{cab}\, \gamma^{ab}_-\eta.
\eeq
Using eqs. (\ref{gamma}) and (\ref{gen}) and the corresponding 
$\eta_\a = c \delta_{\a8}$, this is easily seen to coincide with the 
right-hand side of (\ref{w2}), which is then verified as well as 
weak $G_2$ holonomy. 

As in the previous section, we can also prove the converse: for every weak 
$G_2$ holonomy manifold, we can find a spin connection $\o^{ab}$ verifying 
eqs. (\ref{w1}). Indeed, weak $G_2$ holonomy is equivalent to the existence 
of a spinor $\eta$ satisfying (\ref{w2}). By a suitable local $SO(7)$ 
rotation, we can choose $\eta$ such that $\gamma_+^{ab}\eta =0$ and 
$\partial_i\eta=0$, as in section 3.2. 
Within our representation of the gamma-matrices, which we will use to 
simplify the argument, this means $\eta_\a =c \delta_{\a8}$ 
after the $SO(7)$ transformation. But then (\ref{w2}) reduces to
\beq
\label{w4}
{1\over4}\o_{-j}^{ab}\gamma_-^{ab}\eta = i{\lambda\over8}\gamma_j\eta
\eeq
and, using (\ref{gamma}) and (\ref{gen}), 
we get $\o_-^{ab}\psi_{abc} = -{\lambda\over2}e^c$ which is nothing 
but the first eq. (\ref{w1}). 
We conclude that

\noindent
$\bullet$
{\it The generalized self-duality (\ref{w1}) of the spin connection 
$\o^{ab}$ implies weak
$G_2$ holonomy
and conversely, every weak $G_2$ holonomy metric can be derived 
from a self-dual connection satisfying (\ref{w1}).}

Let us now show how the self-duality condition (\ref{w1}) directly leads 
to $d\phi = \lambda*\phi$, which is the way weak $G_2$ holonomy is usually
stated in the mathematical literature. The three-form $\phi$ and the 
four-form $*\phi$ are still defined as in eq. (\ref{txii}) and $\o^{ab}$
still obeys (\ref{txiii}). Of course, eq. (\ref{txiv}) is now modified:
\beq
\label{w5}
\d\phi = -{1\over2}\psi_{abc}\, e^a e^b \o^{cd} e^d 
= -{1\over4} \psi_{abc}\ph_{cdef} e^a e^b \o^{ef} e^d
+{\l\over 8} \p_{abc}\p_{cde} e^a e^b e^e e^d
=-2\d\phi + 3\lambda*\phi. 
\eeq
where we used (\ref{ident1}) in the last step. So we indeed conclude 
that
\beq
\label{w6}
\d\phi=\lambda*\phi\,.
\eeq

Next we want to investigate the self-duality properties of $R^{ab}$ that 
follow from eq. (\ref{w1}). Clearly, $R^{ab}$ cannot be self-dual as 
it is not Ricci flat. However, eq. (\ref{dxi}) or (\ref{dxii}) suggests
that the two-form
\beq
\label{w7}
\hat R^{ab} = R^{ab} - {\lambda^2\over16}e^a e^b
\eeq
might be self-dual, {\it i.e.} $\psi_{abc}\hat R^{bc}=0$.
This is indeed the case as we now show. Using (\ref{w1}) and 
(\ref{ident3}), we have:
\be\label{w8}
\psi_{abc}\o^{bd}\o^{dc} = \psi_{cab}\left({1\over2}\hat\psi_{bdef}\o^{ef}
-{\lambda\over4}\psi_{bde}e^e\right)\o^{dc} 
= -\psi_{abc}\o^{bd}\o^{dc} - \lambda \o^{ab}e^b 
+{\lambda^2\over8}\psi_{abc}e^be^c \,,
\ee
or
\beq
\label{w9}
\psi_{abc}\o^{bd}\o^{dc} = -{\lambda\over2}\o^{ab}e^b +{\lambda^2\over16}
\psi_{abc}e^be^c \,.
\eeq
Also,
\beq
\label{w10}
\p_{abc} d\o^{bc} = -{\lambda\over2} \d e^a = {\lambda\over2}\o^{ab} e^b \ ,
\eeq
so that indeed
\beq
\label{w11}
\psi_{abc}\hat R^{bc} \equiv \psi_{abc}\left(R^{bc} -
{\lambda^2\over16}e^be^c\right) = 0 \ .
\eeq
Equivalently we have
\beq
\label{w12}
\hat R^{ab} = {1\over2}\hat\psi_{abcd}\hat R^{cd} \ ,
\eeq
where the components of $\hat R^{ab}$ are
$\hat R^{ab}_{\ \ cd} = R^{ab}_{\ \ cd} - {\lambda^2\over16}\bigl(
\delta^a_c\delta^b_d - \delta^a_d\delta^b_c \bigr)$. 
This immediately implies as in eq. (\ref{txix})
\beq
\label{w13}
0 = \hat R_{acbc} = {\cal R}_{ab}- {3\over8}\lambda^2\delta_{ab} 
\quad \Leftrightarrow \quad
{\cal R}_{ij} = {3\over8} \lambda^2 g_{ij}\ ,
\eeq
so that the manifold is Einstein, in agreement with eq. (\ref{dxiii}).

\subsection{Connection with ${\rm spin}(7)$ holonomy}

It is known that a given metric with weak holonomy $G_2$, respectively 
${\rm SU}(3)$, can be associated to a metric with ${\rm spin}(7)$, 
respectively $G_2$, holonomy \cite{Hitchin}. We will show that 
the self-duality (\ref{w1}) of the connection $\o^{ab}$ on the weak 
$G_2$ manifold induces a self-dual $\o^{ab}$ on the ${\rm spin}(7)$ manifold 
so that our results have a natural translation to the latter case as well. 

Let $e^a$ and $\o^{ab}$ be the seven-bein and connection and $\tilde e^A$ 
and $\tilde \o^{AB}$, $A,B=1,\ldots,8$, be the eight-dimensional ones. 
We identify
\beq
\label{w14}
\tilde e^8 = \d t \,, \qquad\qquad \tilde e^a = -{\lambda\over4}t e^a.
\eeq
Then, $\d_{(8)} = {\partial\over\partial t}\d t + \d$, where $d$ is just the 7-dimensional exterior derivative, so that
$\d_{(8)}\tilde e^A + \tilde\o^{AB}\tilde e^B = 0$ implies
\beq
\label{w15}
\tilde \o^{ab} = \o^{ab} \,, \qquad\qquad 
\tilde\o^{8a} = {\lambda\over4} e^a.
\eeq
Define as usual
\beq
\label{w16}
\Psi_{abcd} = \hat\psi_{abcd} \,, \qquad\qquad
\Psi_{abc8} = \psi_{abc}. 
\eeq
The eight-dimensional manifold has ${\rm spin}(7)$ holonomy if and only if
\beq
\label{w18}
\d_{(8)}\Phi = 0 \,, \qquad\qquad
\Phi = {1\over4!} \Psi_{ABCD}\tilde e^A\tilde e^B\tilde e^C\tilde e^D\,.
\eeq
Inserting the ansatz (\ref{w14}), this is
\ba
\label{w19}
\Phi &=& \left({\lambda t\over4}\right)^3\, \d t\wedge \f + 
\left({\lambda t\over4}\right)^4\, *\f \,, \cr
\label{w20}
\d_{(8)}\Phi &=& - \left({\lambda t\over4}\right)^3\, \d t \wedge
\bigl(\d\f-\lambda*\f\bigr) 
+ \left({\lambda t\over4}\right)^4\, \d *\f \,,
\ea
where $\f$ and $*\f$ are the usual seven-dimensional 
3- and 4-form constructed from the $e^a$.
We see that ${\rm spin}(7)$ holonomy for the 
``cohomogeneity-one" ansatz (\ref{w14}) is equivalent 
to weak $G_2$ holonomy of the seven-dimensional level surfaces $t=$const. 
We have shown that for the latter, one can always find a spin connection 
satisfying the self-duality relations (\ref{w1}). In fact, in view of 
eqs. (\ref{w15}), the latter are exactly equivalent to the following ${\rm spin}(7)$ self-duality condition
\beq
\label{w21}
\tilde\o^{AB} = {1\over2}\Psi_{ABCD}\tilde \o^{CD}.
\eeq
This prompts the question whether {\it any} ${\rm spin}(7)$ holonomy metric 
can be derived from (\ref{w21}). Based on the preceeding remarks, one 
would expect this to be true for cohomo\-ge\-nei\-ty-one ${\rm spin}(7)$ metrics. 

More generally, the same proof as for $G_2$ holonomy should work: 
using $\Psi_{ABCD}$ one can again define projectors ${\cal P}_{21}$ 
and ${\cal P}_7$ corresponding to the embedding of 
${\rm spin}(7) \subset {\rm SO}(8)$. Call again $\tilde\o_+^{AB}$, resp
$\tilde\o_-^{AB}$ the part annihilated by  ${\cal P}_7$, 
resp ${\cal P}_{21}$.
The self-duality condition (\ref{w21}) states that $\tilde\o_-^{AB}=0$ 
which is what we want to show. So assume ${\rm spin}(7)$ holonomy.
Then proceeding as in section 3.2, since ${\rm spin}(7)$ holonomy 
implies the existence of a covariantly constant spinor, one shows 
that, after an appropriate local ${\rm SO}(8)$ transformation, 
$D_J \eta =0, J=1, \ldots, 8$, reduces to $\tilde\o_-^{AB} \g^{AB}_-\eta=0$ 
which then implies $\tilde\o_-^{AB}=0$, {\it i.e.} the self-duality equation 
(\ref{w21}).

\subsection{Examples}

\def\d{\delta}
A standard example for weak $G_2$ holonomy is the metric with 
principal orbits given by the Aloff-Walach spaces, see {\it e.g.} 
\cite{KY,CMS}.
It is easy to verify that the standard choice of spin connection satisfies the generalized self-duality (\ref{w1}). 
Further examples and classifications can be found in \cite{FKMS,CS}.

Here, we will consider the example of an eight-dimensional cohomogeneity-one 
${\rm spin}(7)$ metric
with principal orbits being triaxially squashed $S^3$'s over $S^4$ as 
first found by \cite{Cvetic1}. We will show that the differential 
equations imposed by ${\rm spin}(7)$ holonomy on the functions in 
the metric ansatz directly follow from the 
self-duality conditions (\ref{w21}). Reducing to seven dimensions in the way
just described 
the equations for weak $G_2$ holonomy  are just the 
generalized self-duality conditions (\ref{w1}).
 
The starting point is a slight generalization of the ansatz 
of \cite{Cvetic1}:
\beq\label{w22}
ds^2 = dt^2 +a_i^2 R_i^2 + a_\hi^2 P_i^2 + a_7^2 P_0^2\ .
\eeq
Again, in this subsection, $i,j$ run over 1, 2 and 3 only, and 
$\hi = i+3$.
Ref. \cite{Cvetic1} specialises to the case $a_i=a_\hi=a_7$ for 
reasons that will be clear soon. The $R_i$, $P_i$ and $P_0$ as well 
as an additional set of three $L_i$ are left-invariant one-forms of 
${\rm SO}(5)$, and the ansatz (\ref{w22}) corresponds to the coset 
${\rm SO}(5)/{\rm SU}(2)_L$. We define the eight-beins as
\beq\label{w23}
e^a = a_a E_a,\qq e^8 = dt \qq {\rm with} \qq
E_i=R_i\ , \qq E_\hi= P_i\ ,\qq  E_7= P_0 \ .
\eeq
The Maurer-Cartan equations of  ${\rm SO}(5)$ for the 
generators in the coset ${\rm SO}(5)/{\rm SU}(2)_L$ yield 
({\it cf.} \cite{Cvetic1})
\ba\label{w24}
dE_i & = & 
-\e_{ijk} E_j\wedge E_k -\ha E_7 \wedge E_\hi -{1\ov 4} \e_{ijk} E_\hj
\wedge E_\hk\ ,
\nonumber\\
dE_\hi & = & E_7 \wedge E_i - \e_{ijk} E_j \wedge E_\hk 
 + E_7 \wedge L_i + \e_{ijk} L_j\wedge E_\hk\ ,
\nonumber\\
dE_7 & = & E_i\wedge E_\hi  + L_i \wedge E_\hi \ .
\ea 
This leads to
\ba\label{w25}
de^i & = & {\dot a_i \ov a_i} e^8 \wedge e^i -\e_{ijk} {a_i\ov a_j a_k} e^j
\wedge e^k -\ha {a_i \ov a_7 a_\hi} e^7 \wedge e^\hi -{1\ov 4} \e_{ijk} 
{a_i\ov a_\hj a_\hk} e^\hj\wedge e^\hk\ ,
\nonumber\\
de^\hi & = & {\dot a_\hi \ov a_\hi} e^8 \wedge e^\hi 
+{a_\hi \ov a_7 a_i} e^7\wedge e^i 
-\e_{ijk} {a_\hi\ov a_j a_\hk} e^j\wedge e^\hk 
+ {a_\hi\ov a_7} e^7 \wedge L_i + \e_{ijk} {a_\hi\ov a_\hk} L_j \wedge 
e^\hk \  ,
\nonumber\\
de^7 & = & {\dot a_7\ov a_7} e^8\wedge e^7 +{a_7\ov a_i a_\hi} e^i
\wedge e^\hi  + {a_7 \ov a_\hi} L_i \wedge e^\hi \ ,
\\
de^8 & = & 0\ .
\nonumber
\ea
One cannot obtain a torsion-free spin connection 
(satisfying $\o^{ab}=-\o^{ba}$) from these equations 
in general, due to the terms involving the $L_i$. Indeed, one easily 
sees that the terms involving $L_i$ are 
$\o^{7\hi}\vert_{L-{\rm part}}=-(a_7/ a_\hi) L_i$ 
and $\o^{\hi 7}\vert_{L-{\rm part}}=(a_\hi/ a_7) L_i$ 
as well as  
$\o^{\hi\hj}\vert_{L-{\rm part}}= (a_\hi/ a_\hj)\e_{ijk} L_k$.
So to be able to define a torsion-free $\o^{ab}=-\o^{ba}$ we must set
\be\label{w26}
a_\hi=a_7\ .
\ee
Then we obtain for the spin connection
%\ba
%\om^{8a}& = & -{\dot a_a\ov a_a} e^a \ ,
%\nonumber\\
%\om^{ij} & = & -\e_{ijk} \left({a_i\ov a_j a_k} + {a_j\ov a_i a_k} 
%-{a_k\ov a_i a_j}\right)e^k\ ,
%\nonumber\\
%\om^{\hi \hj} & = & -\ha\e_{ijk} \left({a_\hi\ov a_\hj a_k} + {a_\hj\ov 
%a_\hi a_k} -\ha {a_k\ov a_\hi a_\hj}\right)e^k\ ,
%\nonumber\\
%\om^{i \hj} & = & -\ha\e_{ijk} \left({a_\hj\ov a_i a_\hk} - {a_\hk\ov 
%a_i a_\hj }+ \ha {a_i\ov a_\hj a_\hk} \right)e^\hk + {\d_{ij} \ov 2} 
%\left( {a_7\ov a_i a_\hi}- {a_\hi\ov a_7 a_i} -\ha {a_i\ov a_7
%a_\hi}\right) e^7\ ,
%\nonumber\\
%\om^{i7} & = & -\ha \left({a_7\ov a_ia_\hi} - {a_\hi\ov a_7 a_i}+\ha
%{a_i \ov a_7 a_\hi}  \right)e^\hi \ ,
%\\
%\om^{\hi7} & = & -\ha \left(-{a_7\ov a_ia_\hi} - {a_\hi\ov a_7 a_i}+\ha
%{a_i \ov a_7 a_\hi}  \right)e^i \ .
%\nonumber
%\ea
%
\ba\label{w26a}
\om^{8a}& = & -{\dot a_a\ov a_a} e^a \ ,
\nonumber\\
\om^{ij} & = &-\e_{ijk} \left({a_i\ov a_j a_k} + {a_j\ov a_i a_k} 
-{a_k\ov a_i a_j}\right)e^k\ ,
\nonumber\\
\om^{\hi \hj} & = & - \e_{ijk} \left({1\ov a_k} 
-{1\over 4}{a_k\ov a_7^2}\right)e^k +\e_{ijk} L_k\ ,
\nonumber\\
\om^{i \hj} & = &- {1\over 4} \e_{ijk} {a_i\ov a_7^2} e^\hk 
+ {1\over 4}\d_{ij} {a_i\ov a_7^2} e^7\ ,
\nonumber\\
\om^{i7} & = &- {1\over 4} {a_i \ov a_7^2}  e^\hi \ ,
\\
\om^{\hi7} & = &\left({1\ov a_i}-{1\over 4} {a_i \ov a_7^2}  \right)e^i 
+ L_i \ .
\nonumber
\ea
Now we impose the eight-dimensional self-duality conditions 
(\ref{w21}), which read with the present notation
\ba\label{w27}
\om^{8i} & = &-\ha \e_{ijk} (\om^{jk} -\om^{\hj\hk}) + \om^{7\hi}\ ,
\nonumber\\
\om^{8\hi} & = & \e_{ijk} \om^{j\hk} -\om^{7i}\ ,
\nonumber\\
\om^{87} & = & -\om^{i\hi}\ .
\ea
This yields
\beq\label{w28}
\dot a_1 = {a_1^2 -(a_2-a_3)^2\ov a_2 a_3}  -{ a_1^2 \ov 2 a_7^2} 
\ ,\qq ({\rm and\ cyclic})\ , \qq  
\dot a_7 = {1\over4} \sum_{i=1}^3 {a_i\ov a_7} \ .
\eeq
These are exactly the equation found in ref \cite{Cvetic1} for 
${\rm spin}(7)$ holonomy by a very different method.

If we let 
\be\label{w29}
a_i(t)= -{\l\over 4} t A_i  \ , \qq a_7(t) = -{\l\over 4} t A_7 \ ,
\ee
then we know from the previous subsection that the corresponding 
seven-metric 
has weak $G_2$ holonomy, ${\rm d}\f = \l *\f$, provided the constants 
$A_i$ and $A_7$ satisfy the algebraic relations
\be\label{w30}
\l A_1 =-4 {A_1^2 -(A_2-A_3)^2\ov A_2 A_3}  + {2 A_1^2 \ov  A_7^2} 
\ ,\qq ({\rm and\ cyclic})\ , \qq  
\l A_7^2 = - \sum_{i=1}^3 A_i \ .
\ee

\section{Embedding in eleven-dimensional supergravity}\label{sec11dsugra}
\setcounter{equation}{0}

The background geometries considered in the previous sections are thought
to be relevant to $M$-theory. It is then a natural task to consider their
embedding in eleven-dimensional supergravity. 
The bosonic part of the Lagrangian density is \cite{CJS}:
\beq
\label{sugr1}
\begin{array}{rcl}
{\cal L}_{bos.} &=& {1\over2\kappa^2}\biggl[ eR
-{e\over48}F_{MNPQ}F^{MNPQ} \crbig
&& -{1\over3!4!4!}\epsilon^{M_1\ldots M_{11}}
F_{M_1\ldots M_4}F_{M_5\ldots M_8}A_{M_9M_{10}M_{11}}
\biggr] \,,
\end{array}
\eeq
with $F_{MNPQ} = 24\,\partial_{[M}A_{NPQ]}$ and 
$R=R^{AB}_{MN}e^M_A e^N_B$. We take the space-time signature to be
$(-1,1,\ldots,1)$.
We consider this theory on $M_4\times M_7$ with the eleven-bein
\beq
\label{11bein}
e_M^A (X^N) = \left(\begin{array}{cc}
e_\mu^m(x^\nu) & 0 \\ 0 &  e^a_i(x^j) \end{array}
\right).
\eeq
In addition, we introduce bosonic backgrounds
\beq\label{FMNPQ}
F_{\mu\nu\rho\sigma} = f\, e_4\, \epsilon_{\mu\nu\rho\sigma},
\qquad\qquad
F_{ijkl} = \tilde g\,  e_i^a e_j^b e_k^c e_l^d\, 
\hpsi^{abcd},
\eeq
with two constants $f$ and $\tilde g$ and $e_4= {\rm det}\, e_\mu^m$. 
While the first component
is clearly invariant under $O(1,3)\times O(7)$, the invariance of the
second is $O(1,3)\times H$, $H\subset G_2$, depending on the choice of the
seven-bein $ e_i^a$.
With this background, Einstein equations reduce to
\beq
\label{fe1}
R_{\mu\nu} = -{1\over3}\Bigl(f^2+{7\over2}\tilde g^2\Bigr)g_{\mu\nu}\,,
\qquad\qquad
 R_{ij} = {1\over6}\Bigl(f^2+5\tilde g^2\Bigr) g_{ij} \,.
\eeq
The four-dimensional space-time is an Einstein space with negative curvature
whenever $F_{MNPQ}$ has a non-trivial background. 
The field equation for $F_{MNPQ}$ reduces to, 
\beq
\label{fe2}
\tilde g\, (d\phi - f *\phi) =0 \,,
\eeq
where $\phi$ is the three-form 
\beq
\label{phifromF}
\begin{array}{rcl}
\phi &=& {1\over3!}\phi_{ijk}\,dx^i\wedge dx^j\wedge dx^k 
\crbig
&=&{1\over3!4!} e_7\, \epsilon_{ijklmnp}F^{lmnp}
\,dx^i\wedge dx^j\wedge dx^k
\,\,=\,\, {1\over3!} e_i^a e_j^b e^c_k
\psi^{abc}\,\,dx^i\wedge dx^j\wedge dx^k ,
\end{array}
\eeq
with $ e_7={\rm det}\,  e_i^a$. 
Notice at this stage that the appearance of $\phi$ is a mere consequence 
of the $G_2$ structure, while field equation (\ref{fe2}) only
is non-empty if $\tilde g \ne 0$, {\it i.e.} if $*\phi$ is identified 
with a background for $F_{MNPQ}$
on $M_7$. Of course, no information 
on $\phi$ is obtained if $\tilde g=0$, but weak holonomy is required 
by field equations if both $f$ and $\tilde g$ are not zero. 

Supersymmetry breaking with a non-trivial background
for $F_{MNPQ}$ has been studied long ago \cite{BEdWN, CR}. The analysis
is based upon the reduction on the background defined by eqs. 
(\ref{11bein}) and (\ref{FMNPQ}) of the gravitino supersymmetry variation
\beq
\label{deltapsimu1}
\delta\psi_M = {\cal D}_M\epsilon -{1\over288}F_{PQRS}
\left({\Gamma^{PQRS}}_M-8\Gamma^{QRS}\delta^P_M\right)\epsilon \,,
\eeq
omitting all fermionic contributions which vanish on the background.
This variation can be rewritten under $O(1,3)\times O(7)$ using
the following decomposition of the gravitino field and gamma matrices
$\Gamma^M$, $M=(\mu,i)$:
$$
\begin{array}{rcll}
\psi_M &\quad\longrightarrow\quad&
\psi_\mu\otimes\hat\psi_\alpha \,, \qquad&
\psi\otimes\hat\psi_{i,\alpha} \,, 
\crbig
\Gamma^M &\quad\longrightarrow\quad&
\gamma^\mu\otimes I \,, \qquad &
\gamma_5\otimes\gamma^i ,
\end{array}
$$
where $\hat\psi_\alpha$ is a (real) $O(7)$ spinor, $\alpha=1,\ldots,8$,
and $\hat\psi_{i,\alpha}$ is a vector-spinor. One obtains
\ba
\nonumber
\delta(\psi_\mu\otimes\hat\psi_\alpha) &=&
{\cal D}_\mu(\epsilon\otimes\hat\epsilon_\alpha) 
+{1\over12}\tilde g(\gamma_\mu\epsilon)\otimes(\delta_{\alpha\beta}
-8\delta_{\alpha8}\delta_{\beta8})\hat\epsilon_\beta 
-{i\over6}f(\gamma_\mu\gamma_5\epsilon)\otimes\hat\epsilon_\alpha
\crbig 
\label{varia1}
&\equiv& \check{\cal D}_\mu(\epsilon\otimes\hat\epsilon_\alpha) \,,
\crbig
\nonumber
\delta(\psi\otimes\hat\psi_{i,\alpha}) &=& 
{\cal D}_i(\epsilon\otimes\hat\epsilon_\alpha)
+{1\over12}(f-i\tilde g\gamma_5)\epsilon\otimes(i\gamma_i\hat\epsilon)_\alpha
+{i\over3}\,\tilde g(\gamma_5\epsilon)\otimes( e_{i\alpha}\hat\epsilon_8
+3\delta_{\alpha8} e_{i\beta}\hat\epsilon_\beta)
\crbig
\label{varia2}
&\equiv& \check{\cal D}_i(\epsilon\otimes\hat\epsilon_\alpha) \,.
\ea
In principle, the first equation generates the supersymmetry variation 
of potential four-dimensional gravitino fields $\psi_\mu$.
Unbroken supersymmetries require 
\beq
\label{cond1}
0= \delta[\check {\cal D}_\mu,\psi_\nu\otimes\hat\psi_\alpha]
= [\check{\cal D}_\mu,\check{\cal D}_\nu]\,\epsilon\otimes\hat\epsilon_\alpha
\eeq
for at least one direction of the $O(7)$ spinor $\hat\epsilon_\alpha$.
With $[{\cal D}_\mu,{\cal D}_\nu]\epsilon
={1\over4}R_{\mu\nu}^{mn}\gamma_{mn}\epsilon$, the condition leads to
\beq
\label{cond2}
{\cal R}_{\mu\nu}\,(\epsilon\otimes\hat\epsilon_\alpha)
=-{1\over3}g_{\mu\nu}\left[ f^2\,(\epsilon\otimes\hat\epsilon_\alpha)
+{1\over4}\tilde g^2\,(\epsilon\otimes\hat\epsilon_\alpha
+48\delta_{\alpha8}\,\epsilon\otimes\hat\epsilon_8)\right] \,.
\eeq
This is clearly compatible with Einstein equations (\ref{fe1})
only if $\tilde g=0$, in which case, as expected from four-dimensional 
supergravity,
$$
\delta(\psi_\mu\otimes\hat\psi_\alpha) = 
{\cal D}_\mu(\epsilon\otimes\hat\epsilon_\alpha)
- {1\over4}\sqrt{|R|\over3}\,
(i\gamma_\mu\gamma_5\epsilon\otimes\hat\epsilon_\alpha)\,,
$$
with $R= R_{\mu\nu}g^{\mu\nu}$. 

Assuming then $\tilde g=0$, the curvature of the compact space becomes
$ R =  g^{ik} R_{ij} = {7\over6}f^2$, and a conformal
Killing spinor in this space verifies
\beq
\label{Killing}
{\cal D}_i \hat\epsilon = -{1\over2}\sqrt{ R\over42}
(i\gamma_i\hat\epsilon)
= - {1\over12}f (i\gamma_i\hat\epsilon)  \,,
\eeq
which is precisely what is required to cancel the supersymmetry variation
(\ref{varia2}). Then, each solution of eq. (\ref{Killing}) 
produces one four-dimensional supersymmetry. Comparing with 
eq. (\ref{dviii}), one infers that with our background 
\beq
\label{dphieom}
d\phi = -{2\over3}f *\phi \, .
\eeq
This result, which follows from the requirement of four-dimensional
supersymmetry, is incompatible with the field equation (\ref{fe2}) if 
$\tilde g\ne0$. This again shows that supersymmetry breaks if 
$F_{ijkl}$ has a non-zero background value \cite{BEdWN,CR,AS}.

If one chooses a seven-bein $ e_i^a$ leading to a metric for 
the seven-sphere,
%and to an $SO(7)$ connection
there are eight solutions
and the four-dimensional theory has $N=8$ supersymmetry if $\tilde g=0$ 
and $N=0$ if $g\ne0$ \cite{FR,E,DP,BEdWN}.

\section{Conclusions}

We have discussed compactifications of eleven-dimensional supergravity to 
four dimensions with background fluxes and rederived the known result 
that any flux in the internal space breaks all supersymmetry. On the 
other hand, a flux in space-time $\sim f \e_{\m\n\r\s}$ is allowed, 
and some supersymmetry remains unbroken provided the internal space 
admits at least one conformal Killing spinor satisfying 
$D_j \eta \sim i f \g_j\eta$. The flux parameter $f$ thus controls 
the internal geometry which must have weak $G_2$ holonomy for $f\ne 0$ 
and holonomy contained in  $G_2$ for $f=0$.

We have shown that weak $G_2$ holonomy is {\it equivalent} to the spin 
connection satisfying a generalized self-duality condition. Our argument 
is straightforward and is based on the equivalence between weak reduced 
holonomy and the existence of a conformal Killing spinor. The same 
argument applies to (non generalized) self-duality of the spin connection 
and holonomy contained in $G_2$. We also indicated how to tranpose the 
proof to ${\rm spin}(7)$. 

We have given some examples of how the self-duality 
conditions greatly simplify determining the first-order 
differential equations that ensure special (weak) holonomy. Of course, 
explicitly solving these differential equations is another story, as 
always. Actually for all known examples the spin connections directly 
satisfy the appropriate self-duality conditions, without any need to 
first perform a local ${\rm SO}(7)$ (resp. ${\rm SO}(8)$) transformation.

We expect the self-duality relations to be a rather powerful tool to 
generate new solutions with (weak) $G_2$ holonomy. This however, is 
left for future work.

\section*{Acknowledgements}

This work has been supported by the Swiss National Science Foundation,
by the European Union RTN program 
HPRN-CT-2000-00131 and by the Swiss Office for Education and Science.

\section{Appendix: $SO(7)$ and $G2$ algebras, conventions and identities}
\renewcommand{\theequation}{A.\arabic{equation}}
\setcounter{equation}{0}

We review some useful relations and identities for octonions. Some 
useful references are \cite{GG,GT,dWN,FN,FN2,DGT,GN}.
The non-associative octonion algebra is given by (we write $o^a$ rather than the more standard $e^a$ to avoid confusion with the 7-beins) 
\beq
\label{octodef}
o^a o^b + o^b o^a = -2 \delta^{ab}I, \qquad
o^a o^b - o^b o^a = 2 \psi_{abc}o^c, \qquad (a,b,c,= 1,\ldots,7). 
\eeq
While the Clifford algebra is invariant under $SO(7)$, the octonion algebra
and the structure constants $\psi_{abc}$ are invariant under $G_2$.
A four-index invariant tensor can then also be defined:
\be\label{psihatdef}
\hpsi_{abcd}= {1\over 3!}\epsilon^{abcdefg}\psi_{efg}.
\ee
In the standard basis, these tensors are
\beq
\begin{array}{l}
\psi_{123} = \psi_{516} = \psi_{624} = \psi_{435} = \psi_{471} 
= \psi_{673} = \psi_{572} = 1, \crbig
\hpsi_{4567} = \hpsi_{2374} = \hpsi_{1357} = \hpsi_{1276} = \hpsi_{2356} 
= \hpsi_{1245} = \hpsi_{1346} =1 \,.
\end{array}
\eeq
All other non-zero components follow from antisymmetry of the above values. 
A number of useful identities involving products of 
these tensors are listed in eqs. (\ref{ident1}--\ref{ident10}) below. 
A compact  way to write the $\p_{abc}$ is to observe that the indices split into 3 groups as $i=1,2,3$, $\ \hi \equiv i+3 = 4,5,6$,  and 7. Then:
\be\label{split} 
\p_{ijk}= \e_{ijk} \ , \quad
\p_{i\hj\hk}=\p_{\hi j \hk}=\p_{\hi\hj k}=-\e_{ijk} \ , \quad 
\p_{7 i \hj}=\delta_{ij} \ .
\ee

The reduction of the $SO(7)$ spinor into $G_2$, ${\bf8} = {\bf7+1}$, is best 
performed using the antisymmetric and imaginary $SO(7)$
gamma matrices
\beq
\label{gamma}
(\gamma^a)_{\alpha\beta} = i(\psi_{a\alpha\beta} 
+\delta_{a\alpha}\delta_{\beta 8}-\delta_{a\beta}\delta_{\alpha 8}), 
\eeq
where $\alpha,\beta=1,\ldots,8$ are indices in the spinor of $SO(7)$
and $\psi_{a\alpha\beta}$ (and $\ph_{ab\a\b}$) vanish if $\alpha$ or $\beta$ is 8.
With this choice of gamma matrices, $SO(7)$ generators
for the spinor representation are real:
\beq
\label{gen}
\begin{array}{rcl}
\Sigma^{ab} &=& {1\over2}\gamma^{ab} \,\,
=\,\, {1\over4}[\gamma^a,\gamma^b]\,,
\crbig
(\gamma^{ab})_{\a\b} &=& \hat\psi_{ab\a\b}
+\psi_{ab\a}\delta_{\b8}-\psi_{ab\b}\delta_{\a8}
+\delta_{a\a}\delta_{b\b}-\delta_{a\b}\delta_{b\a} \,.
\end{array}
\eeq
The minimal spinor $\eta$ is then also real, we may choose
$\overline\eta=\eta^\tau$ and the reduction 
of a Majorana $SO(1,10)$ spinor $\epsilon$ under $SO(1,3)\times SO(7)$ is 
$\epsilon = \epsilon_4\otimes \epsilon_7$, with a Majorana spinor
$\epsilon_4$. 

A $G_2$ transformation with parameters $\o_{ab}$
of a $SO(7)$ spinor reads then
\beq
\label{spintransf}
\delta_{G_2}\eta_\a = {1\over2}\o_{ab}(\Sigma_+^{ab}\eta)_\a
= {1\over2}\o_{ab}\left( {1\over3}\hat\psi_{ab\a\b}\eta_\b
+{2\over3}\delta_{a\a}\eta_b -{2\over3}\delta_{b\a}\eta_a \right),
\eeq
with generators $J_+^{ab}= \Sigma_+^{ab}$ as defined in eq. (\ref{tviii}).
Clearly, $\eta_8$ is invariant, $\delta_{G_2}\eta_8=0$, 
while $\eta_a$, $a=1,\ldots,7$, transforms in representation ${\bf7}$:
\beq
\label{spintransf2}
\delta_{G_2}\eta_a 
= {1\over2}\o_{cd}\left( {1\over3}\hat\psi_{cdab}\eta_b
+{2\over3}\delta_{ca}\eta_d -{2\over3}\delta_{da}\eta_c \right).
\eeq
Similarly, the action of the generators of the coset $SO(7)/G_2$ is
\beq
\label{spintransf3}
\delta_{SO(7)/G_2}\eta_\a 
= {1\over2}\zeta^c\psi_{abc}(\Sigma_-^{ab})_{\a\b}\eta_\b 
= {1\over2}\psi_{\a c\b}\zeta^c\eta_\b
+{3\over2}(\zeta_\a\eta_8 - \delta_{\a8}\eta_c\zeta^c),
\eeq
with $\zeta_8=0$. In particular, if one starts with 
$\eta_\a = C\delta_{\a8}$, then $\delta_{SO(7)/G_2}\eta_a = 
{3\over2}C\zeta_a$, which is reminiscent from 
$SO(7)/G_2 \sim SO(8)/SO(7)$.

\def\d{\delta}
The tensors $\psi_{abc}$ and $\hat\psi_{abcd}$ can be assembled 
into a single object $\Psi^{\alpha\beta\gamma\delta}$, 
with $\a,\b,\ldots = 1, \ldots 8$, as 
in (\ref{w16}), namely
\beq
\Psi_{abc8}= \psi_{abc}\ ,\qq \Psi_{abcd}=\hat\psi_{abcd}\ .
\eeq
Equation (\ref{psihatdef}) translates into the eight-dimensional self-duality of $\Psi_{\a\b\g\delta}$\ :
\beq 
\Psi_{\a\b\g\d}={1\ov 4!}
\e^{\a\b\g\d\zeta\eta\s\kappa}\Psi_{\zeta\eta\s\kappa}\ .
\eeq
The basic identity for the product of two $\Psi$'s is
\be\label{basicPsiid}
\Psi_{\a\b\g\d}\Psi^{\zeta\eta\s\d}  =  
6 \d^{[\zeta}_\a \d^{\eta}_\b \d^{\s]}_\g
-9 \Psi_{[\a\b}{}^{[\zeta\eta} \d^{\s]}_{\g]} \ .
\ee
It can be used to derive most of the following useful identities for 
products of the tensors $\psi_{abc}$ and $\hpsi_{abcd}$:
\ba
\psi_{abe}\psi_{cde} &=& -\hpsi_{abcd}+\delta_{ac}\delta_{bd} - 
\delta_{ad}\delta_{bc} \,, \label{ident1}
\\
\psi_{acd}\psi_{bcd} &=& 6\, \delta_{ab} \,, \label{ident2}
\\
\psi_{abp}\hpsi_{pcde} &=& 3 \psi_{a[cd}\delta_{e]b} 
- 3\psi_{b[cd}\delta_{e]a} \,, \label{ident3}
\\
\hpsi_{abcp}\hpsi_{defp} &=& -3\hpsi_{ab[de}\delta_{f]c}
-2\hpsi_{def[a}\delta_{b]c} -3\psi_{ab[d}\psi_{ef]c}
+ 6\delta^{[d}_a\delta^e_b\delta^{f]}_c \,, \label{ident4}
\\
\hpsi_{abpq}\psi_{pqc} &=& -4\psi_{abc} \,, \label{ident5}
\\
\hpsi_{abpq}\hpsi_{pqcd} &=& -2\hpsi_{abcd} + 4(\delta_{ac}\delta_{bd} 
- \delta_{ad}\delta_{bc}) \,, \label{ident6}
\\
\hpsi_{apqr}\hpsi_{bpqr} &=& 24\delta_{ab} \,, \label{ident7}
\\
\psi_{abp}\psi_{pcq}\psi_{qde} &=& \psi_{abd}\delta_{ce} 
- \psi_{abe}\delta_{cd} - \psi_{ade}\delta_{bc} + \psi_{bde}\delta_{ac} 
\nonumber\\
&&- \psi_{acd}\delta^{be} + \psi_{ace}\delta_{bd} 
+ \psi_{bcd}\delta_{ae} 
- \psi_{bce}\delta_{ad} \,,  \label{ident8}
\\
\label{ident10} \psi_{paq}\psi_{qbs}\psi_{scp} &=& 3\psi_{abc} \, .
\ea

%%%%%%%%%%%%%%%%%%%%%%%%%%%%%%%%%%%%%%%%%%%%%%%%%%%%%%%%%%%%%%%%%%%%

\end{document}